\documentclass[12pt]{iopart} 

\usepackage{epsfig}
\usepackage{graphics}

\usepackage{color}

\newcommand{\EAI}  {Edwards-Anderson }
\newcommand{\SK}  {Sherrington-Kirkpatrick }
\newcommand{\LD}  {Large Deviations }

\begin{document}

\title{Rare events analysis of temperature chaos in the
  Sherrington-Kirkpatrick model.}  \author{Alain Billoire}
\address{Institut de physique th\'{e}orique, CEA Saclay and CNRS,
  91191 Gif-sur-Yvette, France} 
\eads{alain.billoire@cea.fr}
\begin{abstract} We investigate the question of 
temperature chaos in the Sherrington-Kirkpatrick spin glass model,
applying to existing Monte Carlo data a recently proposed rare events
based data analysis method.  Thanks to this new method, temperature
chaos is now observable for this model, even with the limited size
systems that can be currently simulated.
\end{abstract}
\date{ \today}


\maketitle

The phenomenon of temperature chaos for spin glasses is the extreme
sensitivity of the infinite volume equilibrium state to the slightest
change of the temperature.  In a system with $N$ spins, temperature
chaos means that the equilibrium states at two different temperatures
$T_1$ and $T_2$ (in the same quenched disorder sample) becomes
uncorrelated for $N \gg N^*$, with some crossover $N^*$ that diverges
as $T_1-T_2\to 0$. It was first predicted for finite dimensional spin
glasses~\cite{bray:87b,banavar:87,FH,ney-nifle:93} using either
scaling or Migdal-Kadanoff renormalization group type arguments. Some
numerical evidences have been
presented~\cite{ney-nifle:97,ney-nifle:98,krzakala:04,sasaki:05,katzgraber:07,MonGar}
for temperature chaos in the spin glass phase of the low dimensional
Edwards-Anderson Ising model (EAI) by analyzing the finite size
scaling behavior of the average (both thermal and disorder) overlap
between two clones at different temperatures.  The interpretation of
those numerical results have been criticized recently
in~\cite{Seoane}, where it was argued that the so-called chaos length
and chaos exponents $\zeta$ are not related to temperature chaos, and
that temperature chaos is present for rare disorder samples even in
small volumes, stressing the need to consider individual disorder
samples rather than disorder averaged data.  A new rare events based
analysis method was introduced in this paper as the proper method to
analyze temperature chaos numerically in spin glasses, with the
outcome that there is indeed temperature chaos in the 3D EAI model
with binary distributed quenched couplings, but with qualitatively
different characteristics than was thought before.

Concerning the Sherrington-Kirkpatrick (SK)
model~\cite{sherrington:75,kirkpatrick:78}, the infinite range version
of the \EAI model for which the mean field approximation is exact, no
evidence for temperature chaos have been found numerically despite
heroic efforts~\cite{Chaos1,Chaos2}, although the small excess
observed in~\cite{Chaos2} at low $q$ in the overlap probability distribution
$P(q_{T1,T2})$ for the largest system simulated could be interpreted
as the onset of temperature chaos.  It has later been shown
analytically~\cite{rizzo:03,parisi:10} that there is indeed
temperature chaos in mean field spin glasses by computing, using
Parisi replica techniques, the free energy cost paid in order to
constrain two clones, at temperatures $T_1$ and $T_2$ respectively, to
have a given non-zero overlap (namely to be correlated), and finding a
non-zero solution.  However the effect is extremely weak in the case
of the Sherrington-Kirkpatrick model, due to some accidental
cancellations (for earlier analytical work on temperature chaos in the
\SK model see~\cite{kondor:89,kondor:93,rizzo:01}). It has been argued
that this weakness explains the negative results
of~\cite{Chaos1,Chaos2}, and that it is hopeless to observe chaos
numerically for the \SK model. 

Our aim in this letter is to challenge this pessimistic opinion and re
investigate numerically the question of temperature chaos in the \SK
model, using the rare events method of~\cite{Seoane}.  It is indeed
important to observe numerically temperature chaos in the \SK model,
since the current analytical methods are neither straightforward nor
fully rigorous.  We follow closely~\cite{Seoane} to analyze the
probability density function (pdf) of the reduced chaos overlap
$X^J_{T1,T2}$ between two clones at temperatures $T_1$ and $T_2$
respectively, defined as

\begin{equation}
X^J_{T_1,T_2}=\frac{<q^2_{T_1,T_2}>_J}{
(<q^2_{T_1,T_1}>_J<q^2_{T_2,T_2}>_J)^{1/2}} , \ 
\end{equation}

where $J$ is a quenched disorder sample, and $q_{T1,T2}$ the overlap
between two independent spin configurations with the same disorder
sample $J$ (two real replicas aka clones) and temperatures $T_1$ and
$T_2$ respectively. 
Namely we study the fluctuations of $X^J_{T1,T2}$
with respect to the quenched disorder $J$.  
 Clearly $0\leq X^J_{T_1,T_2} \leq 1$. We use the data
of~\cite{Chaos2}
\footnote{We have extended the statistics of~\cite{Chaos2} in order to
  have $1024$ well thermalized disorder samples for each value of
  $N$. The number of parallel tempering sweeps (defined as a parallel
  tempering sweep per se plus a Metropolis sweep) is $10^6$ for
  measurements after $4 \ 10^5$sweeps for equilibration, but for our
  largest systems where these numbers are $2\ 10^6$ and $8 \ 10^5$
  respectively.}  for the \SK model with binary distributed quenched
couplings, and system sizes $N=256,512,\ldots, 4096$.  With the
parallel tempering algorithm used in~\cite{Chaos1,Chaos2} we have data
for many couple of values of $T_1$ and $T_2$ but we concentrate our
analysis on the values $T_1=0.4$ and $T_2=0.6$, in order to compare
with the 3D EAI results of~\cite{Seoane}, indeed on the one hand it
has been argued in~\cite{Katz} that for binary distributed quenched
couplings the value $T=0.4$ for the SK model is equivalent to the
value $T=0.703$ for the 3D EAI model (used in~\cite{Seoane}). On the
other hand the ratio $(T_c-T_1)/(T_c-T_2)$ are the same in both
situations. The value $T_1=0.4$ is anyway the lowest temperature in
our parallel tempering data, and the value $T_2=0.6$ is a good
compromise between accuracy (that decreases dramatically as $T_2$
decreases towards $T_1$) and the need to stay away from the critical
point $T_c=1$.

\begin{figure}[htb]
\centering
\includegraphics*[height=5cm,angle=0]{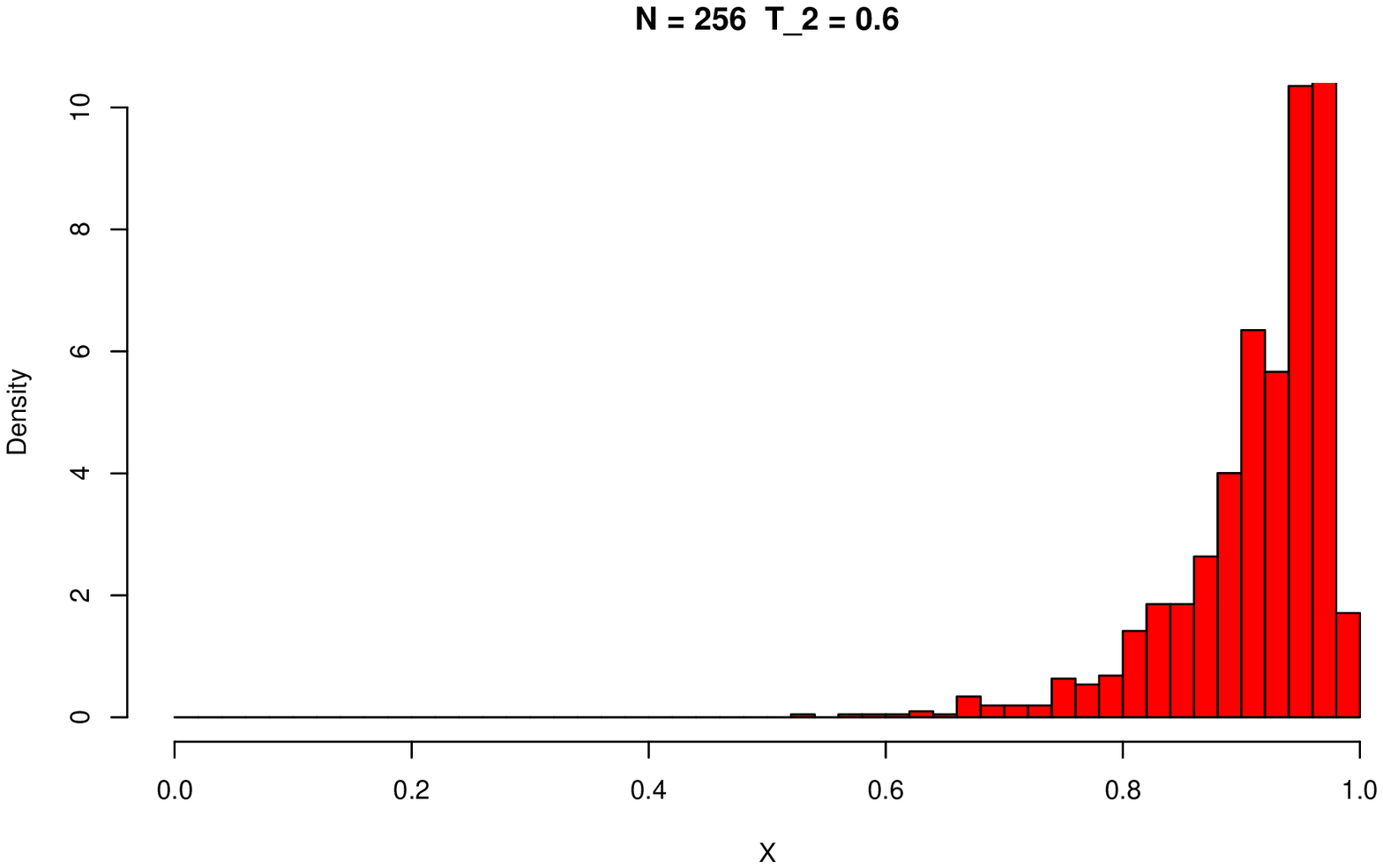}
\includegraphics*[height=5cm,angle=0]{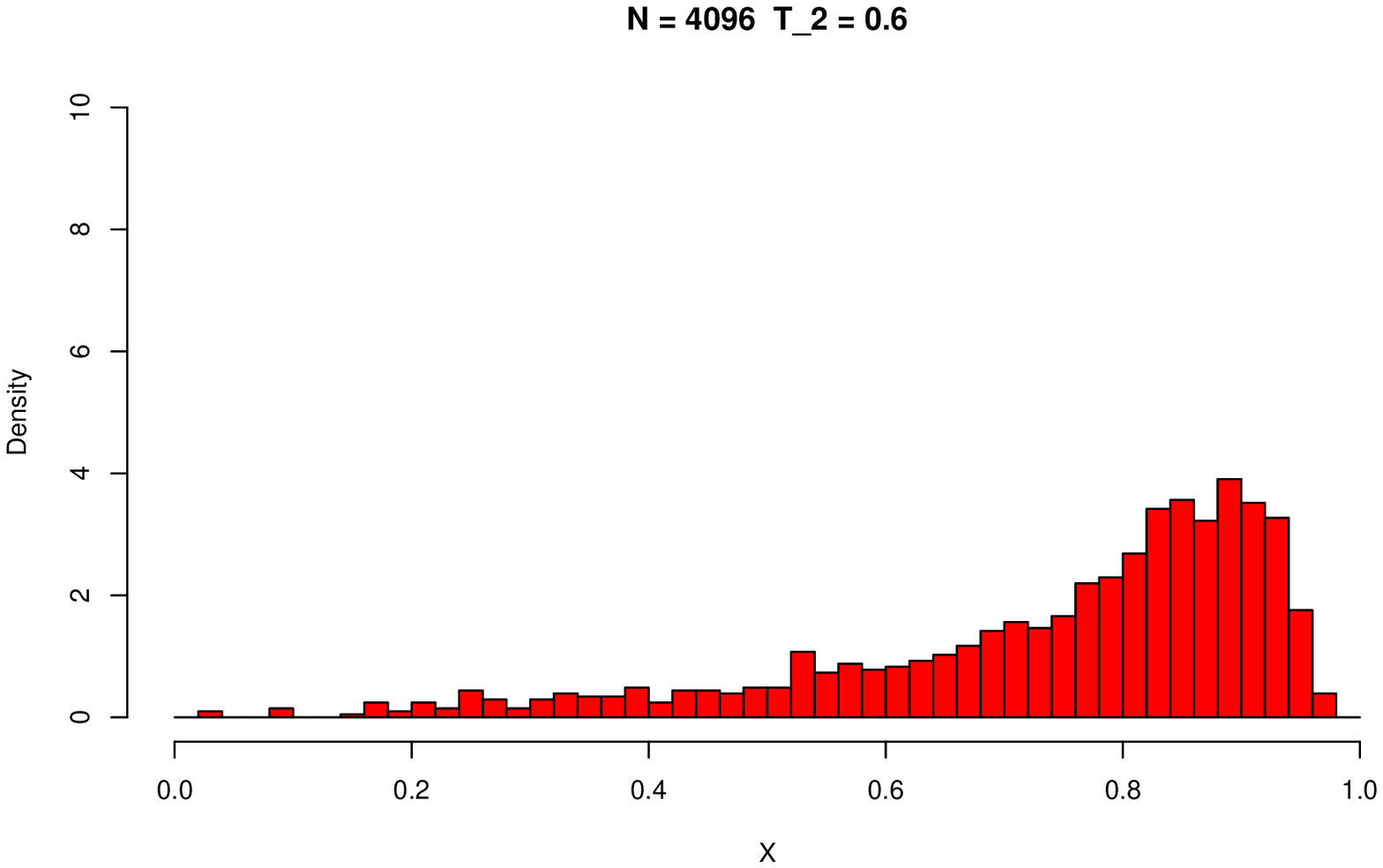}
\caption{Histogram of $X^J_{0.4,0.6}$ for $N=256$ (left) and $4096$
  (right). Both histograms contains data for $1024$ disorder
  samples. The histogram flattens as $N$ grows, with the appearance of
  a low $X$ tail that populates the whole allowed range.}
\label{R}
\end{figure}

In Figure~\ref{R} we show the histogram of $X^J_{0.4,0.6}$ for $N=256$
and $4096$.  If temperature chaos holds, this histogram should
concentrate at the origin ($X^J=0$) for $N\to \infty$. At first glance
the data show just the opposite, with a histogram that is peaked at a
large value $X^J\approx 0.9$ in both cases. There is however a
remarkable broadening of the histogram as $N$ increases from $N=256$
to $4096$, with the appearance of a long low $X^J$ tail. Such a
broadening is quite unusual in statistical physics, and can be
interpreted as the onset of temperature chaos as the following large
deviation analysis shows.

Following~\cite{Seoane} we consider the cumulative distribution
function of the variable $X^J$, and introduce a large deviation (LD)
potential $\Omega^N_{T_1,T_2}(\epsilon)$
\footnote{One should not confuse the   large deviation
  potential $\Omega(\epsilon)$ with the coupled replicas large
  deviation potential $\Delta F(q)$ of~\cite{rizzo:03,parisi:10}. The
  two objects are essentially different, the coupled replicas
  potential describes events that are rare thermodynamically in
  typical samples, while the potential studied here describes
  thermodynamically typical events in rare samples Nevertheless the
  existence of a large deviation potential $\Omega(\epsilon)$ implies
  that typical samples are chaotic as predicted by the non-zero
  $\Delta F(q)$ computed in~\cite{rizzo:03}.}

\begin{equation}
\mbox{Probability}[X^J_{T_1,T_2}<
  \epsilon]=1-\e^{-N\Omega^N_{T_1,T_2}(\epsilon)}\,.
\end{equation}

If this large deviation potential has a non-zero limit as $N\to\infty$
in some temperature interval around $T_1$ (obviously excluding $T_1$
itself), then temperature chaos holds, since $X^J_{T_1,T_2}$ vanishes
in this limit (for any $J$).

\begin{figure}[htb]
\centering
\includegraphics*[height=10cm,angle=270]{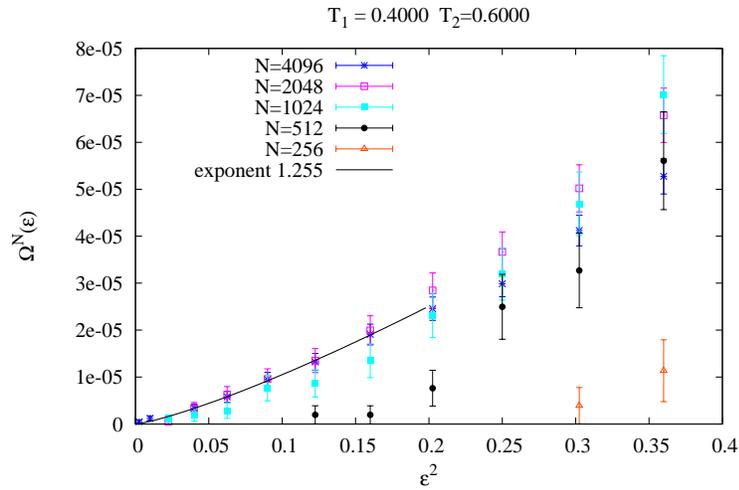}
\caption{The \LD potential $\Omega^N_{0.4,0.6}(\epsilon)$ as a
  function of $\epsilon^2$ for system sizes $N=256, 512,\ldots,
  4096$. We use the standard Wald estimate for the statistical errors.
  Temperature chaos is absent for $N=256$.  For small values of
  $\epsilon$, the potential $\Omega^N_{0.4,0.6}$ increases as $N$
  grows, reaching a limit (within statistical uncertainties) already
  for $N=2048$. The solid line is a fit to the $N=4096$ data. The data
  points with $\epsilon^ 2 \geq 0.2$ are excluded from this fit.}
\label{LD}
\end{figure}

We show in Figure~\ref{LD} the empirical large deviation potential
$\Omega^N_{0.4,0.6}(\epsilon)$, defined as $1/N
\ln(1-\mbox{Probability}[X^J_{0.4,0.6}<\epsilon])$, as a function of
$\epsilon^2$.  For small values of $\epsilon$, the data for our two
larger systems $N=2048$ and $4096$ are compatible, making the case for
a non-zero $N$ independent $\Omega^N_{0.4,0.6}(\epsilon)$ for large
$N$, and consequently for temperature chaos. We note that for small
values of $\epsilon$ the finite size corrections makes
$\Omega^N_{0.4,0.6}(\epsilon)$ smaller, strengthening the case for a
non-zero $N\to\infty$ limit for $\Omega$. (For larger values of
$\epsilon$ however the finite size corrections makes $\Omega$
larger). A fit of the $N=4096$ data for small values of $\epsilon$
shows that $\Omega^N_{0.4,0.6}(\epsilon)\propto \epsilon^{\beta}$ with
$\beta \approx 2.5$. The value $\beta\approx 1.7$ is reported
in~\cite{Seoane} for the 3D EAI model with binary couplings.

\begin{figure}[htb]
\centering
\includegraphics*[height=10cm,angle=270]{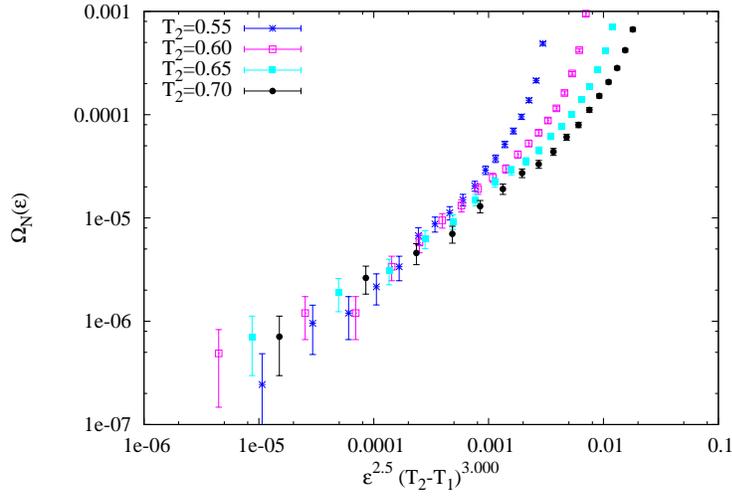}
\caption{Scaling plot (in log-log scale) of the \LD potential
  $\Omega^N_{0.4,T_2}(\epsilon)$ as a function of $T_2-0.4$ for our
  largest system size $N=4096$.  The data scales as a function of
  $y=\epsilon^{\beta} (T_2-T_1)^b$ with $\beta=2.5$ and $b=3$ for
  roughly $y <10^{-3}$. }
\label{Scaling}
\end{figure}

In a finite volume, temperature chaos weakens as $|T_2-T_1|$
decreases, and the onset of chaos is pushed to higher and higher
values of $N$.  It has been suggested~\cite{Seoane} that for small
$\epsilon$ and small temperature difference one has
$\Omega^N_{0.4,0.6}(\epsilon) \propto \epsilon^{\beta} (T_2-T_1)^b$
with $b\approx 1.8$ for the 3D EAI model.  Our data for the SK model
are compatible with such a scaling and a value $b\approx 3$, as show
in Figure~\ref{Scaling} for $N=4096$. Due to large statistical errors
in the small $\epsilon^{\beta} (T_2-T_1)^b$ region where scaling
holds, this is only a rough estimate. An extreme numerical effort
would be needed in order to pinpoint precisely the value of the
exponents $\beta$ and $b$.  A method to either enrich the tail of the
sample distribution corresponding to low values of $q_{T1;T2}$ or to
select the samples belonging to the tail before performing a lengthy
Monte Carlo simulation would be of great help in this matter.

The analytical prediction~\cite{rizzo:03,parisi:10} for the SK model
coupled replicas large deviation potential $\Delta F(q)$ is that
$\beta=7/2$ and $b=2$, but subleading terms are present that cause
transient effects in small volumes.  This may explain the apparent
discrepancy with our numerical results. Another possibility is that
the exponents $\beta$ and $b$ are not the same for the potential
$\Delta F(q)$ and the potential $\Omega(\epsilon)$ studied here.

In conclusion, we have reanalyzed our SK data using the new rare
events based analysis method proposed in~\cite{Seoane}.  We find a
clear signal for temperature chaos in the SK model, at the same
evidence level as the results of~\cite{Seoane} for the 3D EAI model.
In both cases temperature chaos is best seen by analyzing individual
quenched disorder samples: As $N$ grows, chaos first appears with rare
samples that are very chaotic, chaotic samples are more numerous as
$N$ keep on growing, asymptotically, for values of $N$ far beyond
numerical reach, all samples are chaotic.  The finite size scaling of
the distribution of chaotic samples is encoded in a large deviation
potential that scales as a function of overlap squared $\epsilon=q^2$
and temperature difference as $\Omega^N_{0.4,0.6}(\epsilon) \propto
\epsilon^{\beta} (T_2-T_1)^b$, and we give crude estimates of the
values of the exponents $\beta$ and $b$.

\clearpage

\ack The Monte Carlo data used where produced a couple of year ago in
collaboration with Enzo Marinari. I thank him warmly for allowing me
to use these data in this letter.  I acknowledge discussions with
Thomas Garel, Victor Martin-Mayor and C\'ecile Monthus. Numerical
computations have been done at the Bruy\`eres-le-Ch\^atel computer
center.

\end{document}